\documentclass[draft]{agujournal2019}
\usepackage{ragged2e}
\usepackage{url} 
\usepackage{lineno}
\usepackage[inline]{trackchanges} 
\usepackage{soul}
\usepackage{multirow}
\usepackage{enumitem}
\usepackage{amsmath,amssymb,amsfonts}
\usepackage{bm}
\usepackage{booktabs}

\nolinenumbers
\draftfalse

\journalname{JGR: Solid Earth}

\begin{document}
\justifying
%
%


\title{Rock anisotropy promotes hydraulic fracture containment at depth}

%
%




\authors{Guanyi Lu\affil{1}\thanks{Now at University of Pittsburgh}, Seyyedmaalek Momeni\affil{1}, Carlo Peruzzo\affil{1}, Fatima-Ezzahra Moukhtari\affil{1,2}, and Brice Lecampion\affil{1}}

\affiliation{1}{Geo-Energy Laboratory - Gaznat chair, Institute of Civil Engineering, École Polytechnique Fédérale de Lausanne (EPFL), 1015 Lausanne, Switzerland}
\affiliation{2}{Ingphi SA, 1003 Lausanne, Switzerland}




\correspondingauthor{Brice Lecampion}{brice.lecampion@epfl.ch}




\begin{keypoints}
\item Intrinsic anisotropy of sedimentary rocks enhances the horizontal extension and hinders the vertical growth of hydraulic fractures
\item The degree of fracture elongation depends on injection parameters and material properties through a dimensionless toughness coefficient 
\item Elongation is larger for toughness dominated compared to viscous dominated fractures
\end{keypoints}

%
%

%
%


\begin{abstract}
We report laboratory experiments and numerical simulations demonstrating that the anisotropic characteristics of rocks 
play a major role in the elongation of hydraulic fractures propagating in a plane perpendicular to bedding.
Transverse anisotropy leads to 
larger hydraulic fracture extension in the parallel-to-bedding/divider direction compared to the perpendicular-to-bedding/arrester direction. 
This directly promotes vertical containment of hydraulic fractures in most sedimentary basins worldwide even in the absence of any favorable in-situ stress contrasts or other material heterogeneities.
More importantly, the ratio of the energy dissipated in fluid viscous flow in the fracture to the energy dissipated in the creation of new surfaces is found to play a critical role on fracture elongation, with fracture-energy dominated hydraulic fractures being the most elongated while the viscous dominated ones remain more circular. These results open the door to a better engineering and control of hydraulic fractures containment at depth in view of the competition between material anisotropy and injection parameters (fluid viscosity and rate of injection).
\end{abstract}

\section*{Plain Language Summary}
The widespread application of hydraulic fracturing for unconventional hydrocarbon production has prompted concerns about fractures extending vertically to sensitive rock layers, highlighting the need to understand fluid-driven fracturing for informed public discourse and improved industrial practices. Through laboratory experiments and numerical simulations, we show that the intrinsic anisotropic characteristics of sedimentary rocks lead to limited hydraulic fracture height growth across the bedding planes in the most common geological situations. Furthermore, we quantify the roles of elastic constants, fracture toughness, as well as the fluid injection conditions in shaping hydraulic fracture in transversely isotropic rocks. Our findings suggest that the hydraulic fracture is most elongated in the toughness-dominated regime, and the impact of rock anisotropy vanishes when the fracture propagates in the viscosity-dominated regime.

%
%

%


%
%
%
%

\section{Introduction}

Hydraulic fractures are widely used for the production enhancement of wells in unconventional hydro-carbon resources among other applications \cite{Deto16}. These tensile fractures propagate quasi-statically in rocks due to the injection of fluid at pressures greater than the minimum in-situ compressive stress. 
These fractures grow perpendicular to the minimum in-situ stress direction, which is in most sedimentary basins horizontal, such that hydraulic fractures propagate vertically \cite{HuWi57}. Controlling the vertical height growth of a hydraulic fracture has long been considered a key factor for successful applications since the desire is to create a fracture that extends to the full height of the reservoir, while preventing excessive vertical growth that could create communication pathways for unwanted fluid migration into adjacent strata \cite{EcNo00,FiWa12,BuLe17}. Over the years, concerns have been raised over serious environmental issues, such as contamination of underground drinking water resources by upward migration of fracturing fluid \cite{HoIn11,OsVe11,WaJa12,ViBr13,VeJa14,EPA16} and compromised seal integrity of the caprock in geologic carbon sequestration \cite{Schr07,FuSe17}, both of which may happen as a result of unbounded vertical fracture growth. Therefore, it is essential to accurately predict and control the vertical propagation of hydraulic fractures. However, predicting fracture height is particularly challenging, as numerous field evidences suggest that the actual height of a hydraulic fracture often differs from what is predicted by state of the art hydraulic fracturing models \cite{SmMo15}. Microseismic and tiltmeter monitoring data from thousands of hydraulic fracturing treatments indicates that the induced fractures are generally more constrained in the vertical direction and are longer laterally compared to theoretical predictions \cite{FiWa12,FlTy13}.

The limitations of the vertical growth of hydraulic fractures are traditionally thought to be a result of strong variation of in-situ stresses and  material properties across rock formations \cite{SiAB78,WaSc82,vanE82,JeBu09spej,XiYo18}, as well as interaction with pre-existing discontinuities in/across different rock formations \cite{TeCl84,WaTe87,ZhJe07,ZhCh08}. However, hydraulic fractures more elongated horizontally than vertically have also been observed in homogeneous formations not exhibiting any increase in confining stress vertically that could explain this limited height growth \cite{CiCo18,KoZo21}. We argue that - rock anisotropy - an intrinsic characteristic of sedimentary rocks has a first order impact on the shape of hydraulic fractures and thus their ultimate vertical extent. 
Unconventional hydrocarbon reservoirs are formed primarily in sedimentary basins, which have strong anisotropic material properties at a fine scale thanks to their deposition and diagenesis \cite{HoSc94,SoZo13}. More specifically, the anisotropy is caused by a common directional feature of sedimentary rocks – beds, which are generally sub-horizontal planes formed during the deposition of the sediments. Mechanical properties of sedimentary rocks, such as mudstones and shales, are found to vary substantially along different directions with respect to the bedding planes \cite{HeGu15,ZhSh18,Mouk20,LuDu21}. They are widely modeled as a transversely isotropic material at the continuum scale \cite{JoWa81,JoCh95,Wang02b,MoLe20,LuDu21}. A recent theoretical study has shown that the shape of a vertical hydraulic fracture that grows perpendicular to the bedding direction in a transversely isotropic material differs remarkably from what would be expected in an isotropic medium, indicating a strong impact by the rock's anisotropic characteristics on the vertical containment of hydraulic fractures at depth \cite{MoLe20}. 

In the following, we bring together laboratory hydraulic fracturing experiments and numerical simulations to uncover the key factors that govern the vertical containment of hydraulic fractures at depth in a transversely isotropic rock formation. It is important to re-inforce that we are here mostly interested in the propagation of planar hydraulic fractures in a plane perpendicular to bedding - a configuration of most practical relevance at depth - and do not address propagation in the bedding plane (which is typically favored at shallow depth / low confining stresses).

\section{Extensive acoustic monitoring methods to capture laboratory hydraulic fracture evolution}

\begin{figure*}[!ht]
\begin{centering}
\includegraphics[width=.9\columnwidth]{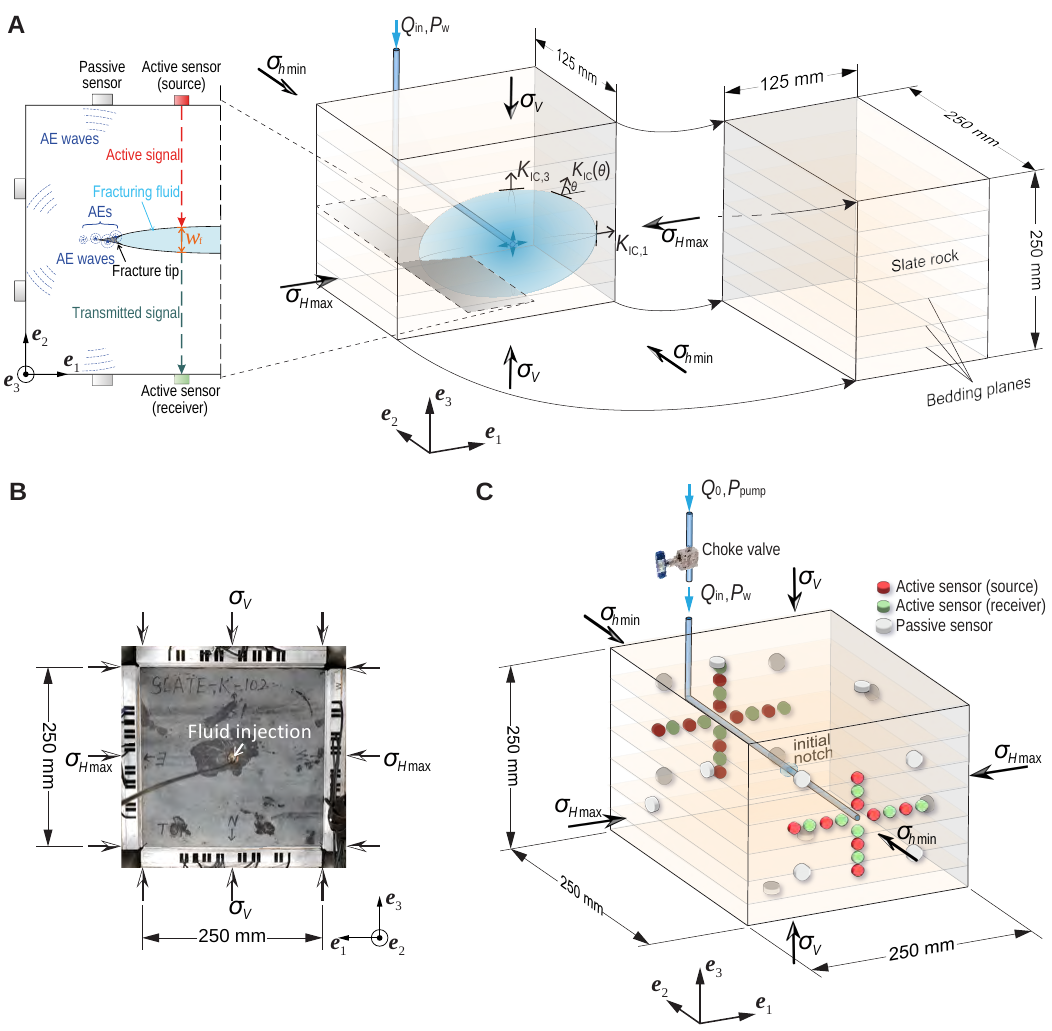}
\par\end{centering}
\caption{A, Cross sectional view of a propagating hydraulic fracture in the experiments on an anisotropic slate (right) with a conceptual sketch of the passive (acoustic emissions) and active (wave transmission) acoustic monitoring system (left). The experiments are designed to mimic a vertical hydraulic fracture propagating in a layered rock formation at depth. B, Photograph of the Carmen slate block under confinements. C, Active and passive acoustic sensor layout. \label{fig:1}}
\end{figure*}

A total of four hydraulic fracturing experiments are carried out on cubic blocks of Del Carmen slate in a true-triaxial load frame (Figure \ref*{fig:1}). Del Carmen slate is a finely laminated metamorphic rock with an extremely small porosity from La Ba\~na, Le\'on, North-West Spain. It exhibits two typical transversely isotropic properties: (1) an anisotropic variation of the critical energy to propagate a fracture as function of the fracture growth direction with respect to the plane of isotropy (bedding plane in transversely isotropic rocks), $K_\text{IC}(\theta)$ (Figure \ref*{fig:1}A), and (2)  five independent elastic constants $C_\text{ij}$ (or the corresponding anisotropic elastic moduli $E^\prime(\theta)$ \cite{Cher12,LaUl14,MoLe20} and Thomsen parameters \cite{thom86}). These properties have been measured in laboratory \cite{Mouk20} and are reported in \ref{appendix_A}. 

The slate specimens used in the experiments are 250$\times$250$\times$250-mm cubic blocks (Figure \ref*{fig:1}B). All samples are prepared with an axisymmetric notch (10-mm radius) that emanates from the center of the horizontal wellbore with 8-mm radius. The fracture is driven by the injection of a Newtonian fluid in the axisymmetric notch through a wellbore drilled in the center of the specimen. The injection system is separated into two parts by a choke valve: (1) the upstream that starts from the pump and ends before the valve (under constant pumping rate, $Q_0$, and pump pressure, $P_\text{pump}$), and (2) the downstream that consists of the fluid passing through the valve and flowing into the fracture (with a fluid influx of $Q_\text{in}$ and wellbore pressure of $P_\text{w}$). Three types of fluid are used as the injection fluid (Table \ref*{tab:1}): (1) Mixture of glycerol and water is used in K1 to facilitate fracture growth in a toughness-dominated regime, (2) T2 uses $99\%$ glycerol for maintaining the propagation in a transition regime, and (3) glucose is used in M3 and M4 to target for the viscosity-dominated regime hydraulic fracturing growth. The bedding plane is set to be orthogonal to the fracture plane to replicate the in-situ condition of a vertical hydraulic fracture growth at depth in sedimentary basins (Figure \ref*{fig:1}A). We apply a sufficiently large vertical stress (normal to the bedding plane) in a true triaxial frame with the following confining stresses for all four tests: $\sigma_v$ = 20 MPa, $\sigma_{H\text{max}}$ = 13 MPa, $\sigma_{h\text{min}}$ = 0.5 MPa. This setup maximizes the vertical extent of the created fracture and avoid any deviation of the fracture into a bedding plane.

\begin{table*}[!ht]
\caption{Summary of testing conditions.}\label{tab:1}
\centering{}%
\begin{center}
\begin{tabular}{ l l l l l }
\toprule 
\multirow{2}{*}{Test ID} & $Q_0$ & $\mu$ & Test duration & \multirow{2}{*}{Regime} \\
& (ml/min) & (Pa$\cdot$s) & $\textit{T}_\text{exp}$ (\text{s}) &  \\
\midrule
K1  & 0.04  & 0.11  & 82  & Toughness-dominated \\
T2  & 0.08  & 0.617 & 55  & Transition \\
M3  & 0.15  & 25.4  & 558 & Viscosity-dominated \\
M4  & 0.15  & 25.4  & 508 & Viscosity-dominated \\
\bottomrule
\end{tabular}
\begin{tablenotes}
\footnotesize
\end{tablenotes}
\end{center}
\end{table*}

Extensive acoustic measurements, via both passive and active acoustic methods, are used to image the hydraulic fracture propagation. The appearance of micro-cracks adjacent to the macro-scale fracture is accompanied by the emission of transient elastic waves due to the release of strain energy, which is referred to as acoustic emissions (AEs) \cite{Lock93,ShLa95,ChLe04,HaGu18,HaGu19,LuDu21}. Our passive acoustic monitoring network consists of 16 piezoelectric sensors mounted on all six surfaces of the block as shown in Figure \ref*{fig:1}C. Throughout the experiments, each of the 16 VS150-M Vallen resonant (at 150 KHz) piezoelectric sensors, covering frequencies from 100 KHz to 1 MHz, records AEs in a continuous mode with a sampling rate of 10 MHz. The 3D hypocenter location of the AE events are obtained by a semi-automatic algorithm using a modified Time Difference Of Arrival (TDOA) method \cite{Kund14,MoLi21}, with the compressional-wave velocities of the rock at different orientations measured for intact specimens. The relative magnitudes of the AEs are estimated based on wave amplitudes and source-to-receiver distance \cite{ZaWa98}. 

In parallel to passive monitoring, an active acoustic array consisting of 16 source-receiver sensor pairs allows us to track the evolution of the macro-scale fracture. These source-receiver pairs, mounted on two opposite vertical faces parallel to the hydraulic fracture plane (Figure \ref*{fig:1}A and C), enable estimation of the fracture width at 16 locations via an analysis of transmitted waves with a 90$^\circ$ incident angle. In a three-layer geometry (rock-fluid-rock) as shown in Figure \ref*{fig:1}A, the thickness of the fluid layer (i.e., fracture width), $w_\text{f}$, is evaluated by matching the spectrum of the transmitted signals travelling between two facing source-receiver transducers with the predicted values \cite{GrFo98,LiLe20,LiLe22a}. More specifically, we solve for $w_\text{f}$ by minimizing the difference between the transmitted signal and the product of a reference signal and a transmission coefficient in the frequency domain \cite{GrFo98,LiLe20}. Repetitive acoustic surveys are carried out at a fixed time interval (every 10 seconds), using a total of 32 Controltech resonant (at 750 KHz) piezoelectric compressional-wave transducers (16 source-receiver pairs) with frequency coverage from 100 KHz to 4 MHz. Each survey consists of 50 source excitations using the Ricker function with a peak frequency of 750 KHz that are stacked to improve the signal-to-noise ratio, and $w_\text{f}$ is computed at every survey.

The simultaneous passive and active acoustic monitoring provide a wealth of information on both micro-fracturing and macro-scale hydraulic fracture width evolution. Integrating these methods allows to successfully capture the three-dimensional (3D) evolution of hydraulic fracture growth in these experiments.

\section{Fracture elongation induced by material anisotropy}

Hydraulic fracture growth in low permeability isotropic material is well known to be governed by the relative influence of the energy dissipated in fracture creation with respect to the energy dissipated in viscous fluid flow in the fracture \cite{SaDe02,Deto04,BuDe08}. It results in two distinctly different regimes of propagation: a toughness-dominated regime where the energy spent in fracture creation dominates, and a viscosity-dominated regime where fluid viscous energy dissipation dominates. Our aim is to study the impact of transverse isotropy on the overall shape of hydraulic fracture propagating in both types of propagation regimes.

\begin{figure*}[!ht]
\begin{centering}
\includegraphics[width=\columnwidth]{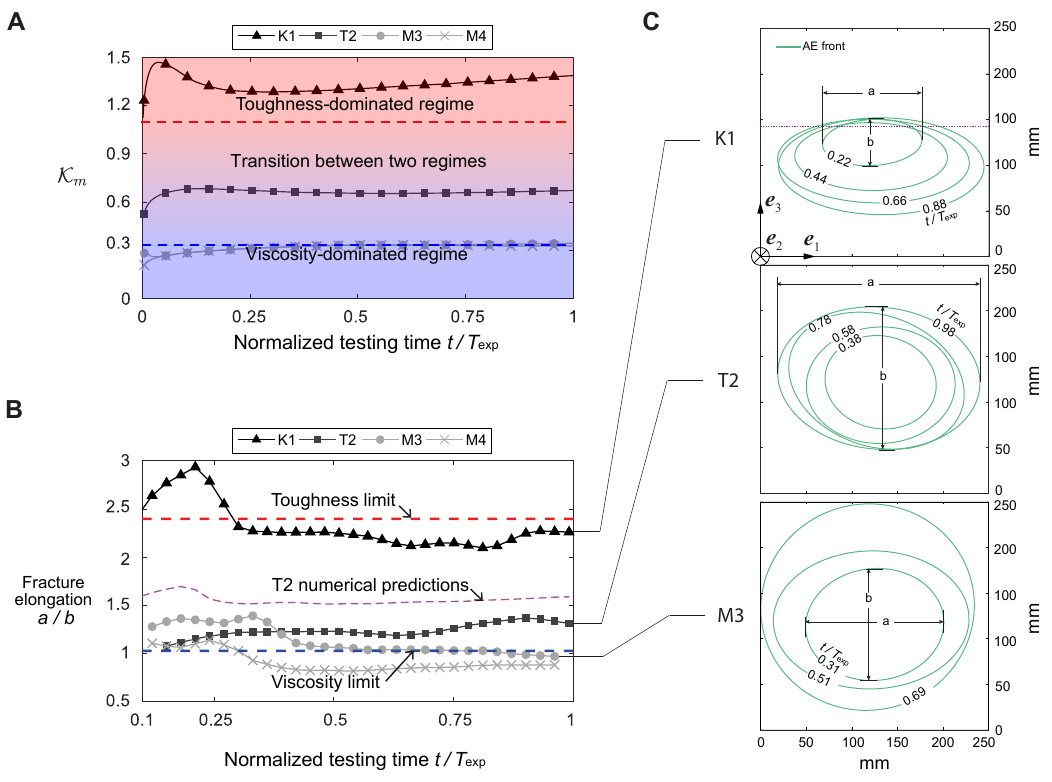}
\par\end{centering}
\caption{A, Evolution of the dimensionless toughness $\mathcal{K}_m$ with normalized testing time, $t/T_\text{exp}$, where $T_\text{exp}$ is defined as the total test duration in each experiment. Here, $E^\prime$ and $K_\text{IC}$ are computed as the average of their two extreme values along $\bm{e}_1$ and $\bm{e}_3$ directions, $E^\prime=(E_{1}^\prime+E_{3}^\prime)/2$, $K_\text{IC}=(K_\text{IC,1}+K_\text{IC,3})/2$, with $E_{1}^\prime=E^\prime(\theta=0)$, $E_{3}^\prime=E^\prime(\theta=\frac{\pi}{2})$, $K_\text{IC,1}=K_\text{IC}(\theta=0)$, and $K_\text{IC,3}=K_\text{IC}(\theta=\frac{\pi}{2})$. B, $a/b$ measured in all tests plotted together with reference values in limiting cases in both toughness- and viscosity-dominated regimes and simulation results of the transition test. C, Elliptical shape of the hydraulic fractures reconstructed using AE data in the toughness-dominated (K1), transition (T2), and viscosity-dominated (M3) experiments. The most elliptical shape is observed in the toughness-dominated regime, and a radial hydraulic fracture (no elongation) is seen in the viscosity-dominated test. \label{fig:2}}
\end{figure*}

\begin{figure}
\begin{centering}
\includegraphics[width=\textwidth]{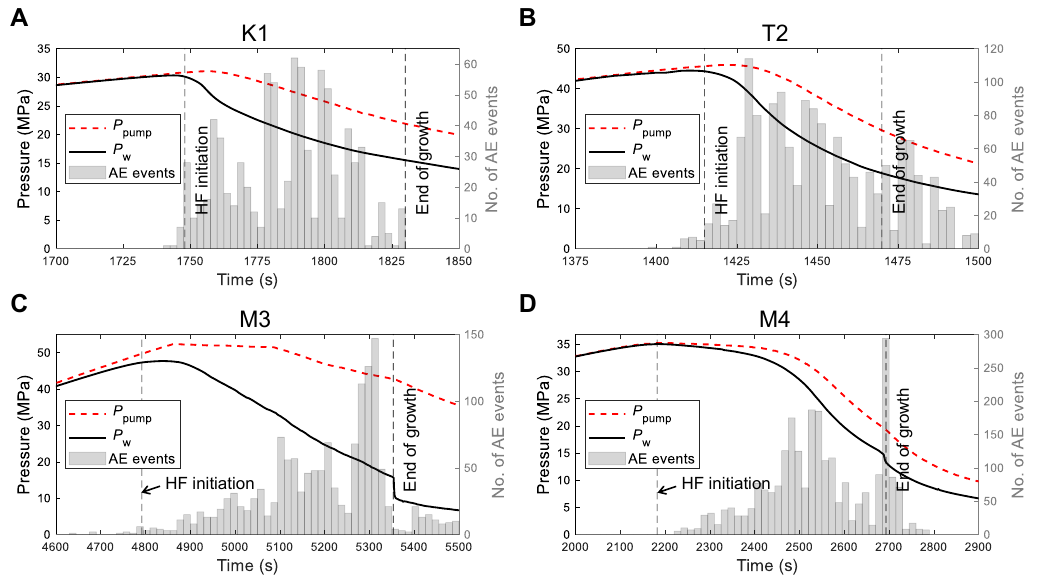}
\par\end{centering}
\caption{Evolution of pump pressure, $P_\text{pump}$, wellbore pressure, $P_\text{w}$, and number of AEs with time of all experiments. The time of hydraulic fracture (HF) initiation and end of growth are determined by clear signs such as pressure change, and the location and number of AEs. In both M3 and M4, very few AEs were detected in the first $\sim$100 seconds after the fracture initiation (initiation time determined by change in slope of the wellbore pressure as fluid starts to flow into the hydraulic fracture). Therefore, this period of time was disregarded in the analysis of AE front evolution. \label{fig:3}}
\end{figure}

The relative influence of these two dissipative mechanisms (surfaces creation and viscous flow) on hydraulic fracture growth can be quantified by a dimensionless fracture toughness $\mathcal{K}_m$ obtained from scaling considerations \cite{SaDe02,Deto04,BuDe07jem,HuGa10,LuGo17,LeDe17}. In fact, it is computed as the square root of the ratio of fracture creation to viscous flow energy dissipation. Accounting for a time-varying injection from a point source, it is given by 
\begin{equation}
\mathcal{K}_m(t) = \frac{{K_\text{IC}t^{5/18}}}{{E^\prime}^{13/18}{\mu^\prime}^{5/18}V_\text{in}(t)^{1/6}} \label{eq:dimensionless_K}
\end{equation}
where $K_\text{IC}$ is the fracture toughness, $E^\prime=E/(1-\nu^2)$ represents the plane strain elastic modulus, $\mu^\prime=12\mu$ with $\mu$ the dynamic viscosity of the injection fluid, and $t = T-T_0$ where \textit{T} and $T_0$ are the absolute and fracture initiation time, respectively. $V_\text{in}(t)=\int_{0}^tQ_\text{in}(\tau)\text{d}\tau$ represents the total volume of fluid in the fracture, where $Q_\text{in}$ is the fluid influx into the fracture accounting for wellbore compressibility \cite{LiLe22a}.
A radial hydraulic fracture in an isotropic material \cite{SaDe02} grows in the toughness-dominated regime for $\mathcal{K}_m \geq 1.1$, and in the viscosity-dominated regime when $\mathcal{K}_m \leq 0.32$. 
We consider any value ranging from 0.32 to 1.1 as a transitional regime between the two limits. Similar scaling laws hold for a transversely isotropic rock \cite{MoLe20} pending the use of an average characteristic value for the toughness and elastic modulus. 
The experimental conditions are summarized in Table \ref*{tab:1}. By varying the fluid injection conditions (fracturing fluid viscosity and injection rate), we aim for specific propagation regimes (toughness-dominated, viscosity-dominated, and transition regimes) in different experiments. Figure \ref*{fig:2}A gives the evolution of $\mathcal{K}_m$ with normalized testing time, $t/T_\text{exp}$. Two experiments were performed under viscosity-dominated regime (hereafter denoted as M3 and M4), experiment K1 was in toughness-dominated regime, and T2 is considered to be in the transition regime. The pressure and AE events histories in all four experiments are plotted in Figure \ref*{fig:3}. 

\begin{figure*}[!ht]
\begin{centering}
\includegraphics[width=\columnwidth]{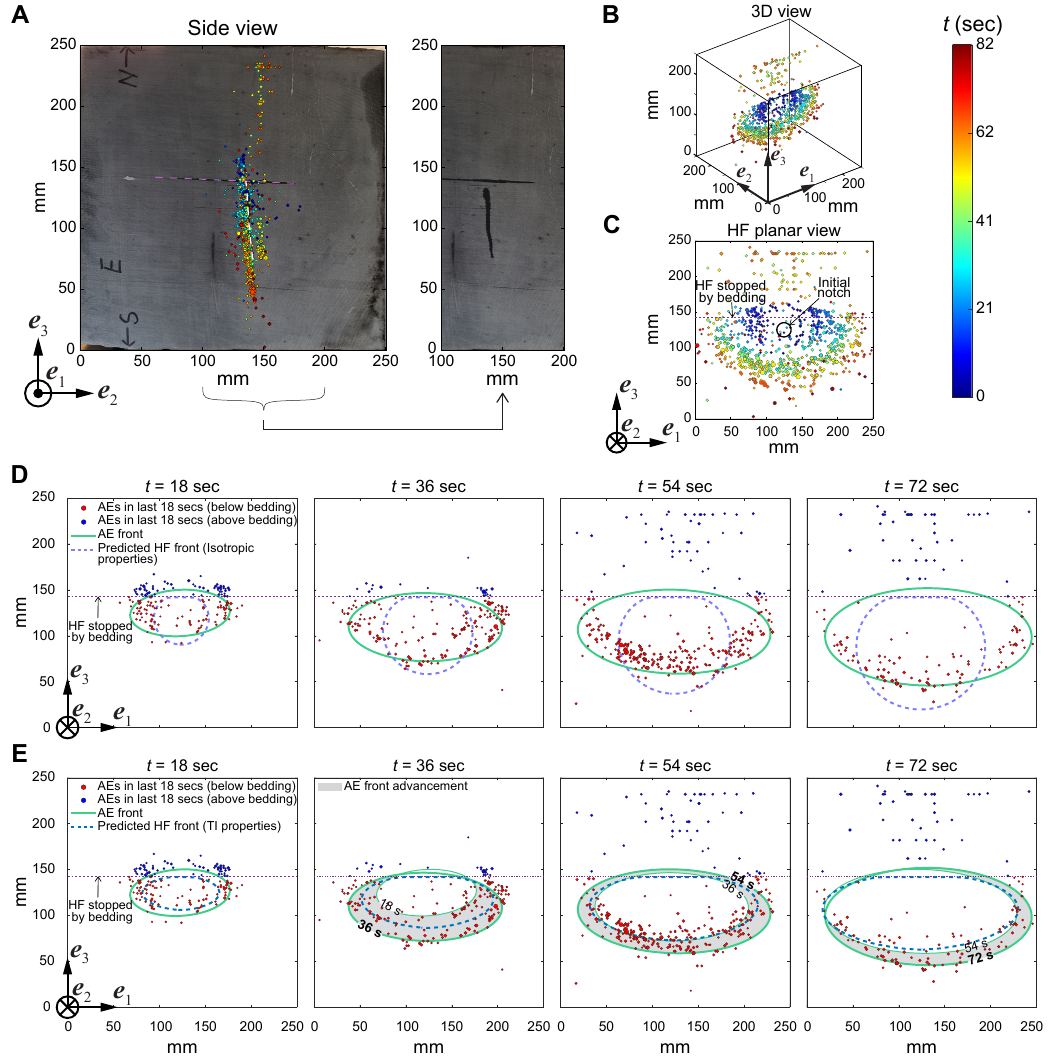}
\par\end{centering}
\caption{Experimental and simulation results of test K1. A, Hypocenter location of the AEs plotted on $\bm{e}_2\bm{e}_3$-plane, superimposed on the post-test photograph of the sample surface, with the event occurrence time and magnitude indicated by the color and size of the circles, respectively. The final hydraulic fracture (HF) plane (white dashed line) is seen to be completely stopped by a visible bedding plane (purple). B, C, 3D and $\bm{e}_1\bm{e}_3-$planar view of the event hypocenter locations. D, Four snapshots taken in different times demonstrating the comparisons between the reconstructed AE front and predicted fracture front assuming hydraulic fracture growth in an isotropic medium (projected on the same 250$\times$250-mm $\bm{e}_1\bm{e}_3-$plane as in C). Events that occur within the last $\Delta t=18$ secs before the time of every snapshot are also plotted. In the elliptical front reconstruction for K1, the events located above the bedding plane are disregarded since the hydraulic fracture was stopped by the weak plane, and these events are considered as pure micro-cracks that do not coalesce into the macro-scale fracture. E, Comparisons between AE front and predicted fracture front obtained from the hydraulic fracture solver considering a transversely isotropic medium. The grey area highlights the advancement of the AE front between two snapshots. \label{fig:4}}
\end{figure*}

First, we focus on the toughness-dominated regime test, K1. AE data is collected throughout the experiment. A majority of the events are concentrated along the final fracture plane as the AE hypocenters (Figure \ref*{fig:4}A) overlap with the fracture plane highlighted in the post-test photograph of sample surface (also confirmed by the AE density plot in Figure S1 in Supporting Information S1). The upward growth of the hydraulic fracture along positive $\bm{e}_3$ direction (perpendicular-to-bedding) was stopped by a specific bedding plane located $\sim$2 cm above the wellbore (which was visible on the specimen surfaces before the test when wetted). Although there was no interruption of the hydraulic fracture by any bedding plane below the wellbore, the fracture plane did not reach the bottom face. On the contrary, we observe larger fracture length along $\bm{e}_1$ direction (parallel-to-bedding), as the fracture eventually extended to the full length of 250 mm in $\bm{e}_1$ direction. The event locations projected on the 2D $\bm{e}_1\bm{e}_3$-plane in Figure \ref*{fig:4}C also suggest that micro-cracking extends further along the parallel-to-bedding direction compared to the perpendicular-to-bedding direction.

This finding is consistent with numerical and analytical studies that suggest an ellipse-like shape for a fluid driven fracture propagating in a transversely isotropic rock \cite{LaUl14,BeDo18,Dont19,MoLe20}. Following these previous works, and considering that AEs generally take place in the adjacent areas of the growing fracture front, it is sensible to assume that the frontier formed by the AEs also expands with an elliptical shape. This assumption enables us to reconstruct a generalized AE front by solving a least-squares problem (Text S2 in Supporting Information S1) to fit the best ellipse for the outermost events that occur within a given time interval (green ellipse in Figure \ref*{fig:4}D). Four snapshots of the reconstructed fronts are shown in Figure \ref*{fig:4}D. The major and minor semi-axes are found to be aligned generally with the $\bm{e}_1$ and $\bm{e}_3$ directions, implying a clear elongation of the hydraulic fracture growth along the parallel-to-bedding direction. To investigate the elongation of the hydraulic fracture, we measure the ratio of the fracture extent along $\bm{e}_1$ direction, $a$, over its value along $\bm{e}_3$ direction, $b$. As shown in Figure \ref*{fig:4}C, $a$ increases to as high as $\sim$3 times of $b$, indicating a significant elongation of the hydraulic fracture along the parallel-to-bedding direction.

To demonstrate that such elongated fracture growth is due to the rock's transversely isotropic characteristics, instead of being caused by the specific bedding plane interrupting the fracture propagation, the experimental results are compared with numerical predictions by two models: an isotropic model and a transversely isotropic one for the rock. These simulations are carried out using an extensively verified planar 3D hydraulic fracturing solver \cite{ZiLe20,MoLe20} as detailed in Text S2 in Supporting Information S1. 
The first model solves the problem of a hydraulic fracture propagating in an isotropic medium with a constant fracture toughness \cite{ZiLe20} (independent of the propagation direction). To prevent the fracture from advancing across the observed specific bedding plane, a jump in fracture toughness is imposed at the bedding location, to a level that is much higher than the uniform fracture toughness of the medium. As a result, the fracture initially grows in a radial shape until hitting the bedding (Figure \ref*{fig:4}D). As its propagation is partially disrupted by the bedding plane, its center begins to shift downward in an effort to maintain a somewhat radial shape, and eventually reaches the bottom surface prior to hitting both vertical faces. Substantial discrepancies are found between the numerical predictions and the reconstructed AE fronts. To summarize, for an isotropic rock, the arrest of the propagation of a planar hydraulic fracture on one side would enhance, instead of suppressing, its growth in the opposite direction.
In the second model, we account for the transversely isotropic features of the medium \cite{ZiLe20,MoLe20}. More specifically, the rock's elastic deformation and resistance to creation of new fracture surfaces induced by fluid pressure now depend on five transversely isotropic elastic constants, as well as the anisotropic fracture toughness. Detailed comparisons between the AE fronts and the predicted fracture fronts provided in Figure \ref*{fig:4}E reveal that: (1) The AE front constantly propagates ahead of the hydraulic fracture front; (2) both fronts advance at roughly the same pace; (3) the AE clusters are scattered initially and become more concentrated in the predicted fracture front region. 
The elliptical AE front has a better agreement with the predicted fracture front when accounting for transverse isotropy compared to the radial shape as typically observed in hydraulic fractures in an isotropic medium. We will use the transversely isotropic model as numerical predictions for hydraulic fracture growth hereafter.

\section{Effect of viscous fluid dissipation on fracture elongation}

\begin{figure*}[!ht]
\begin{centering}
\includegraphics[width=\columnwidth]{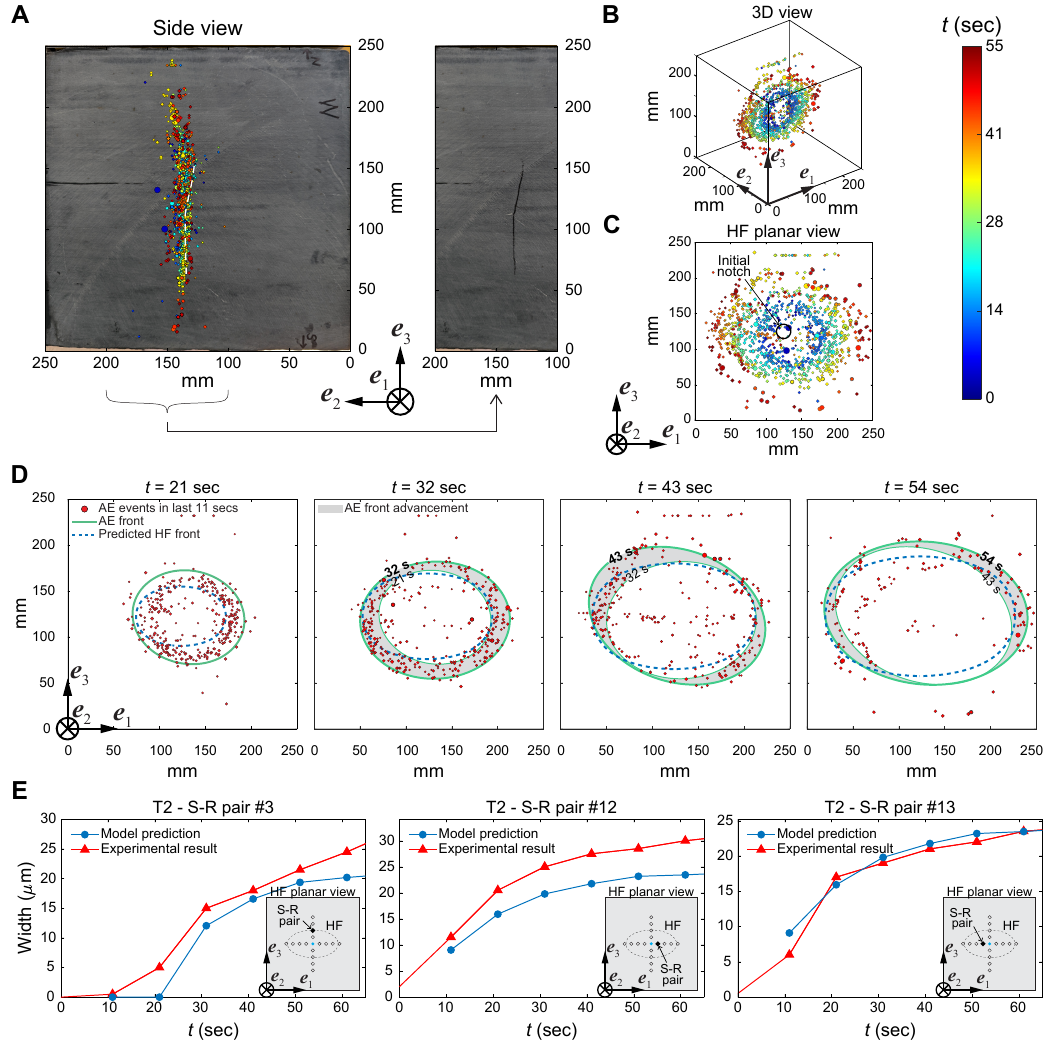}
\par\end{centering}
\caption{Experimental and simulation results for the T2 test (transition regime). A$-$D, Post-test photograh, 3D and $\bm{e}_1\bm{e}_3-$planar view of the event hypocenters compared with model predictions. E, Width evolution at source-receiver (S-R) pairs \#3, 12, and 13 in T2, with sensor locations indicated on the $\bm{e}_1\bm{e}_3$-plane. The width is 5$\sim$10 $\mu\text{m}$ larger than the model prediction. Such discrepancy can be explained by the fact that elastic constants used in numerical modeling are based on ultrasonic wave-speed measurements, which, in general, are higher than their quasi-static values. The numerical solver thus likely underestimates the fracture width due to overestimation of the elastic stiffness constants.\label{fig:5}}
\end{figure*}

\begin{figure*}[!ht]
\begin{centering}
\includegraphics[width=\columnwidth]{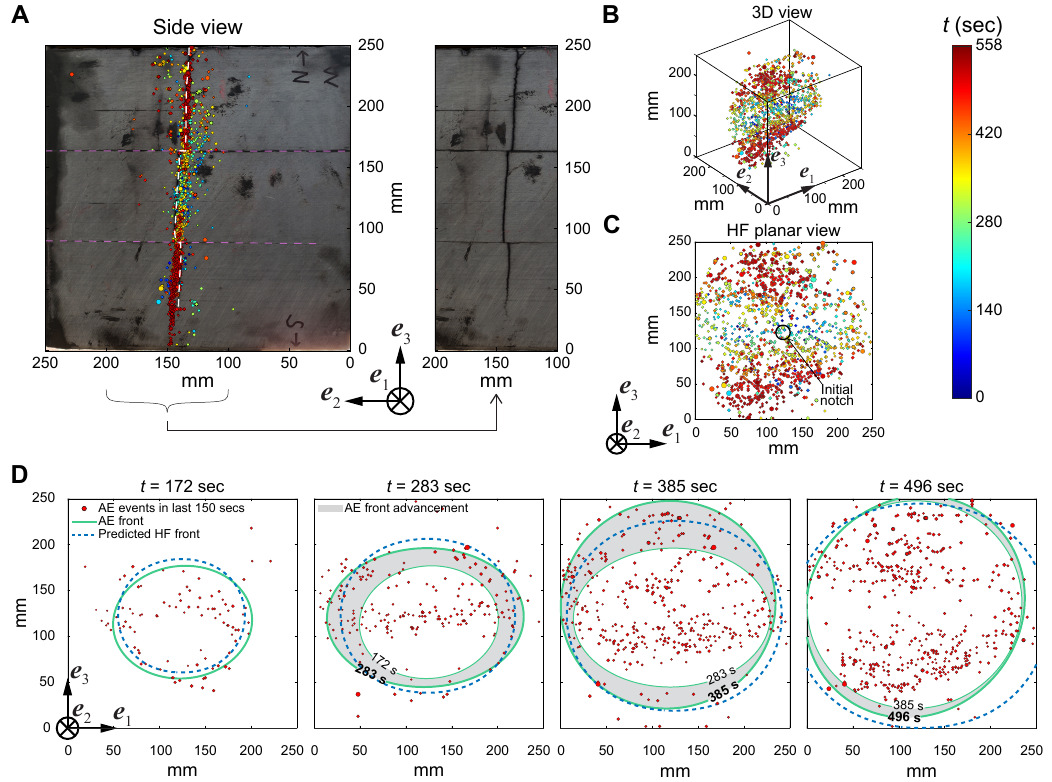}
\par\end{centering}
\caption{Experimental and simulation results of the M3 test (viscosity-dominated regime).
A$-$D, Post-test photographs, 3D and $\bm{e}_1\bm{e}_3-$planar view of the event hypocenters compared with model predictions.
\label{fig:6}}
\end{figure*}

Next, we demonstrate the fracture growth in experiments under other regimes: experimental results for the transition regime (T2) and the viscosity-dominated regime tests (M3 \& M4) are displayed in Figure \ref*{fig:5}, Figure \ref*{fig:6} and Figure S3 in Supporting Information S1, respectively. In all experiments, the final fracture plane remains vertical with little inclination (as illustrated by the AE density plots in Figure S1 in Supporting Information S1). The evolution of the AE frontier in all tests indicates that: (1) the effect of rock anisotropy is most significant in the toughness-dominated regime, which evidently promotes fracture containment in the perpendicular-to-bedding direction; (2) as the propagation regime transitions towards the viscosity-dominated regime (decreasing $\mathcal{K}_m$), the reconstructed front becomes less elliptical, and the AEs are found to be more scattered across the entire fracture plane. Post-test visual examination on the fracture path confirms larger vertical growth in the two viscosity-dominated tests compared to the toughness-dominated and transition tests.

Notably, the predicted fracture width in T2 converges to the measured one at multiple locations near the wellbore (Figure \ref*{fig:5}E), confirming that the hydraulic fracture was centered at the wellbore and remained vertical during the experiment. 
In the viscosity-dominated tests, regardless of the scattering in the events, it is seen that the AE front matches well with the predicted fracture front throughout the lifetime of both specimens. The fracture width in M3 (Figure S4 in Supporting Information S1) increases together with the numerical solution in both early and late times. However, a drop at an intermediate time in most measurement locations (\#10, 11 and 15) is observed. This phenomenon is possibly associated with the occurrence of a fluid lag - as often observed in viscosity-dominated hydraulic fracture tests \cite{BuDe08}. Strong elasto-hydrodynamics coupling in the near-tip region of a hydraulic fracture induces cavitation such that the fluid front lags behind the fracture tip \cite{GaDe00}. Consequently, the acoustic signal cannot travel through the hydraulic fracture when it hits this near-tip nonwetted zone, which leads to erroneous estimations of the fracture opening \cite{LiLe22b}.

Theoretically, the impact of rock anisotropy was found to vary between propagation regimes \cite{MoLe20}. However, for a given regime, the fracture shape evolves in a self-similar manner and can be grasped by the aspect ratio between the major and minor semi-axes of the fracture footprint noted as $a/b$. Two relations have been proposed for $a/b$ corresponding to hydraulic fracture propagation respectively in the toughness- and viscosity-dominated regimes \cite{MoLe20}. The aspect ratio is the lowest in the viscosity-dominated regime, and is found to evolve as 

\[
a/b \approx \left[0.76(E_3^\prime/E_1^\prime)^{1/3}+0.24\right]^{-1}
\]

In the toughness-dominated regime, the elongation is more pronounced and the aspect ratio scales as 

\[a/b\approx{\left(\dfrac{K_\text{IC,3}\cdot E_1^\prime}{K_\text{IC,1}\cdot E_3^\prime}\right)^2}
\]

The values of $a/b$ measured from lab experiments are compared to these two analytical solutions, as well as the numerical predictions (Figure \ref*{fig:2}B and C). We find in K1 that $a/b$ firstly rises above the toughness regime limit when the hydraulic fracture growth is partially stopped by the bedding. Its value then drops and converges to the toughness limit as the hydraulic fracture regains the elliptical shape as its center is shifted downward. The reconstructed AE front in T2 is seen to be less elliptical compared to the model predictions at the beginning as the AEs are more scattered, but it approaches the numerical predictions as the hydraulic fracture propagation continues.

Interestingly, we observe an unexpected elliptical front shape in both viscosity-dominated regime tests at the beginning of propagation, and $a/b$ ultimately decreases to $\sim$1 (approaching a circular footprint). Such initial uneven fracture growth is likely related to the fluid lag at the beginning of fracture growth. It has been established \cite{Gara06b,LeDe07,BuDe07jem} that although the fluid lag may be large at early-time of the propagation, it ultimately coalesces with the fracture front over a characteristic time-scale of order ${E^\prime}^3 \mu^\prime/\sigma_{h\text{min}}^3$ (where $\sigma_{h\text{min}}$ is the minimum confining stress).
In the case of an initially significant fluid lag, the shape of the hydraulic fracture is primarily determined by the effect of the anisotropic fracture toughness (as the tip is dry) and elastic constants, which explains why an elliptical shape is observed at early time, while the fracture becomes ultimately more radial as the fluid reaches the fracture front.

\begin{figure*}[!ht]
\begin{centering}
\includegraphics[width=.7\columnwidth]{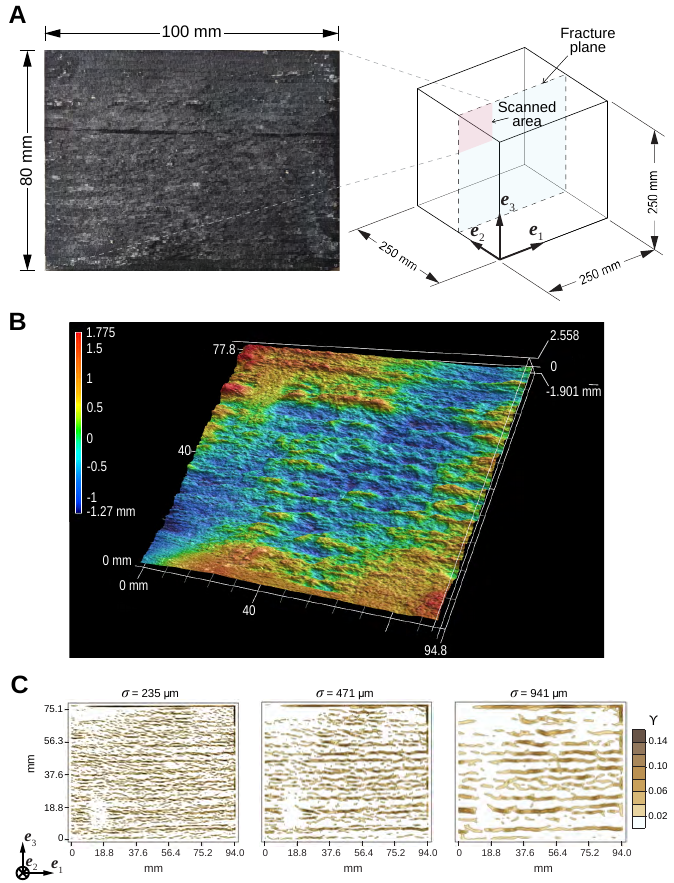}
\par\end{centering}
\caption{A, Post-test photograph of the direction-dependent rough fracture surface of a part of the fracture plane created in M3 (80 mm $\times$ 100 mm). B, 3D profile of the same partial fracture plane mapped by a Keyence VR-3200 optical profilometer with a voxel resolution of 47$\mu\text{m}^3$. C, An estimate for the main principal surface curvature $\Upsilon$ (see Text S3 in Supporting Information S1) of the original surface elevation profile
is plotted at three different scales (from left to right: $\sigma=235\mu m$, $\sigma=471\mu m$, $\sigma=941\mu m$).
The repeated bands of stark contrast representing high curvature are oriented in the same direction as the bedding ($\bm{e}_1$) and coincide with parallel grooves recognisable by direct inspection of the fracture surface.
\label{fig:7}}
\end{figure*}

The intrinsic anisotropy of the rock is also evident in the topography of the created fracture surfaces. 
The post-test photograph and the 3D roughness profile 
of part of the fracture plane created in M3 (Figure \ref*{fig:7}A and B), as well as the main principal surface curvature plots in Figure \ref*{fig:7}C show a clear direction-dependent rough surface characterized by parallel grooves aligned with the orientation of bedding planes. 
The importance of heterogeneity on controlling fracture roughness has been recently well quantified for a model material (hydrogel) \cite{StRu22}.
The anisotropic fracture roughness, with a rougher texture across the bedding and a smoother texture in the parallel-to-bedding direction, can thus be attributed to different length-scales of heterogeneity in the perpendicular and parallel-to-bedding directions associated with the rock deposition.
Naturally, the propagation of fractures along the rougher direction necessitates higher energy consumption compared to the smoother direction, which is thereby speculated to be one factor that causes the elongation of these fractures.

\section{Conclusions}

Clear correlation between the aspect ratio of a hydraulic fracture, $a/b$, and the dimensionless toughness, $\mathcal{K}_m$, is revealed by both laboratory experiments and numerical simulations (Figure \ref*{fig:2}A and B) - larger $\mathcal{K}_m$ results in a more elongated hydraulic fracture shape that restricts the fracture growth in the perpendicular-to-bedding orientation, whereas smaller $\mathcal{K}_m$ leads to a more isotropic propagation. We conclude that rock anisotropy has a dominating effect on the vertical containment/horizontal elongation of hydraulic fractures in the absence of variation in confining stresses and material properties. We have clearly demonstrated that the intrinsic layering of shale formation, which is reflected in their transversely isotropic behavior, can favor the containment of hydraulic fracture at depth when the orientation of the material isotropy plane is perpendicular to the minimum confining stress. Such a configuration is ubiquitous in sedimentary basins worldwide. Including the effect of rock anisotropy more systematically - with proper material characterization - is clearly needed in order to reconcile field observations further. More importantly, our results highlight the fact that the injection parameters (larger fracturing fluid viscosity and larger injection rate) can suppress the beneficial impact of material anisotropy on fracture containment by increasing the energy spent in viscous fluid flow (see Eq.~(\ref{eq:dimensionless_K})). The results presented here open the door to a more consistent engineering design of hydraulic fracturing treatments in shales with respect to their confinement at depth.

%
%
%
%
\appendix
\section{Rock characterization} \label{appendix_A}

Mineral composition and organic content of the Del Carmen slate are determined through X-ray Powder Diffraction analysis \cite{Mouk20}. We observe a concentration of 35.7\% of laminated silicates (in particular chlorite and mica), 43.61\% of quartz, and some minor constituents such as plagioclases (12.84\%) and feldspars (3.15\%). Values of  elastic constants  $C_\text{ij}$ are determined by ultrasonic measurements of compressional- and shear-wave velocities \cite{Tsva12,Wang02a} (Table \ref{tab:2}). The fracture toughness is measured by three-point loading on semicircular bending specimens \cite{KuOb14}. Samples are prepared in two orientations with respect to the bedding planes to measure two extreme values of $K_\text{IC}$ along the parallel-to-bedding ($K_\text{IC,1}$) and perpendicular-to-bedding ($K_\text{IC,3}$) directions, respectively. The direction dependent anisotropic fracture toughness, $K_\text{IC}(\theta)$, used in the numerical predictions is then determined by a specific form \cite{Mouk20} (function of $K_\text{IC,1}$, $K_\text{IC,3}$, $\theta$, and $C_\text{ij}$) that ensures an exact elliptical shape for a planar fracture under uniform loading. The elastic properties appear to be highly consistent among all samples, as the standard deviations of the ultrasonic wave-speeds are, on average, less than 1\% of the mean values (for example, $V_{P}(\theta=0)=6432\pm41$ m/s).

\begin{table*}[!ht]
\caption{Transversely isotropic elastic properties and fracture toughness of the Del Carmen slate rock. The subscripts represent the orientations of the anisotropic properties ($\bm{e}_1$ or $\bm{e}_3$).}\label{tab:2}
\centering{}%
\begin{center}
\begin{tabular}{ l l l l }
\toprule 
Elastic &  Plane-strain & Thomsen  & Fracture  \\
constants &  elastic modulus & parameters & toughness  \\
 $C_\text{ij}$ (GPa) &  (GPa) &    & (\text{MPa}$\sqrt{\text{m}}$) \\
\midrule
$\textit{C}_{11}=114.6$ &  $E_{1}^\prime=107.5$  & $\epsilon=0.26$ & $K_\text{IC,1}=2.5$ \\
$\textit{C}_{12}=28.4$  &  $E_{3}^\prime=97.2$  &  $\gamma=0.1$  & $K_\text{IC,3}=3.5$ \\
$\textit{C}_{13}=4.7$   &    &  $\delta=0.013$  & \\
$\textit{C}_{33}=75.5$  &    &    & \\
$\textit{C}_{44}=35.9$  &    &    & \\
\bottomrule
\end{tabular}
\end{center}
\end{table*}


\section*{Open Research Section}
The raw active acoustic data, as well as the processed experimental data, including the fluid injection records and detailed acoustic emission results are available on Zenodo (\url{https://doi.org/10.5281/zenodo.7738236}). We also provide processed experimental data as source data for the figures used in this paper. 
The raw passive acoustic dataset for the 16 channels recorded in continuous mode  is too large (several TBs) to share in a public repository but can be made available upon request. 

The hydraulic fracture simulator Pyfrac is available at \url{https://github.com/GeoEnergyLab-EPFL/PyFrac}. The source code for fracture width estimation from the active acoustic measurement is available at \url{https://github.com/GeoEnergyLab-EPFL/FracMonitoring.git}.

\acknowledgments
This work was funded by the Swiss National Science Foundation through grants no.~160577 \& no.~192237. We are grateful to Prof. Pedro M. Reis for generously providing access to the surface roughness measurement equipment. We thank Dr. T. Adatte for the X-ray powder diffraction measurement of the Del Carmen slate mineralogy.

\section*{Author Contributions}

Conceptualization: Guanyi Lu, Brice Lecampion

Data curation: Guanyi Lu, Seyyedmaalek Momeni, Carlo Peruzzo

Formal analysis: Guanyi Lu, Seyyedmaalek Momeni, Carlo Peruzzo, Fatima-Ezzahra
Moukhtari, Brice Lecampion

Funding acquisition: Brice Lecampion

Investigation: Guanyi Lu, Seyyedmaalek Momeni, Carlo Peruzzo

Methodology: Guanyi Lu, Seyyedmaalek Momeni, Carlo Peruzzo, Brice Lecampion

Supervision: Brice Lecampion

\section*{Author Declaration}

The authors declare no competing interests.

\pagebreak

\textbf{\large Supporting Information for "Rock anisotropy promotes hydraulic fracture containment at depth"}

\noindent\textbf{Contents of this file}
\begin{enumerate}
\item Text S1 to S4
\item Figures S1 to S4
\end{enumerate}

\noindent\textbf{Introduction}

Text S1 presents an overview of the reconstruction of AE front based on measured AE events.

Text S2 details the numerical model.

Text S3 provides details of measurements and analysis on the surface roughness of the hydraulic fracture.

Text S4 explains the fluid lag observed in the viscosity-dominated tests.

%


\noindent\textbf{Text S1. AE front reconstruction}

To track the advancement of the micro-cracking frontier, we first pick the outermost events (one event from every $10^{\circ}$ angular sector) detected in a fixed time interval prior to the time of interest (150 seconds time window in the example of T2, shown in Figure S2). Next, an ellipse-specific fitting method \cite{Book79,FiPi99} is applied to reconstruct the elliptical acoustic emission front as recovered in Figure S2.

\noindent\textbf{Text S2. Numerical model}

The planar 3D open-source hydraulic fracture simulator - Pyfrac \cite{ZiLe20,MoLe20} - is used to simulate the growth of a fluid driven fracture. It has been extensively validated against both analytical solutions and experiments performed in isotropic materials. 
The fracture is set to propagate on the plane perpendicular to the isotropy plane. The input parameters in the numerical model, including the rock and fluid properties, minimum confining stress and the injection rate history $Q_\text{in}(t)$, are set to be the exact same values as in the laboratory experiments.

\noindent\textbf{Text S3. Fracture surface roughness measurement}

The 3D surface elevation profile of part of the fracture plane created in M3 (see Figure 7) has been measured using an optical profilometer (VR-3200, Keyence Corporation). 
To highlight the strong anisotropy of the surface, a second-order Gaussian derivative filter has been applied at different scales $\sigma$ to estimate the local Hessian matrix. The main negative eigenvalue of the Hessian matrix which corresponds to the main principal curvature orthogonal to a ridge is then plotted in Figure 4 (see 
Mathematica\textsuperscript{\textcopyright} RidgeFilter command for details).

\noindent\textbf{Text S4. Fracture width analysis of viscosity-dominated regime tests}

Unlike the continuously increasing width obtained in the toughness-dominated and transition regime tests, the width evolution in the viscosity-dominated regime tests generally experiences three phases (Figure S4), which may be attributed to the occurrence of fluid lag \cite{GaDe00,BuDe08} - cavitation near the tip region of a hydraulic fracture due to strong elasto-hydrodynamics. These three stages are inferred by the width analysis with active acoustic data (Figure S4B): I. Once the fracture tip arrives at the location that intervenes with the active signal, the two fracture faces start to separate as the width is observed to be increasing with time. During this phase, the fracture faces are still bonded by inter-granular forces, such that an interface, filled with deformed rock particles, is formed. The acoustic signals are still able to travel through this bonded interface. Consequently, we see increase in the fracture width in this rock-solid interface-rock geometry. II. The two fracture faces become fully debonded when its width reaches 20$\sim$30 microns, while the fluid front is still lagging behind. The appearance of the lag zone hinders wave transmission, and results in erroneous values in the width evaluation as huge drops are seen during this phase. III. Finally, the lagging fluid front advances to this region and occupies the previously void space, which reopens the pathway for the acoustic signals to travel through the intermediate fluid layer. Width estimation regains its accuracy since wave transmission is restored in this phase.

\begin{figure}
\begin{centering}
\includegraphics[width=\textwidth]{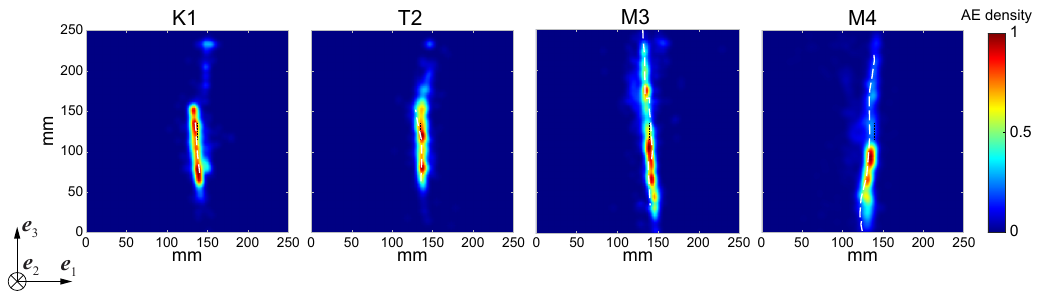}
\par\end{centering}
\caption{Acoustic emission density plots of all tests (side view). The concentration of the events is scaled to the range between 0 and 1 (indicated by different colors), with 0 corresponding to no event and 1 corresponding to the maximum density. In all tests, planar hydraulic fractures are observed with highest concentration of events along the final fracture planes (white dashed lines). \label{fig:AE_density}}
\end{figure}

\begin{figure}
\begin{centering}
\includegraphics[width=\textwidth]{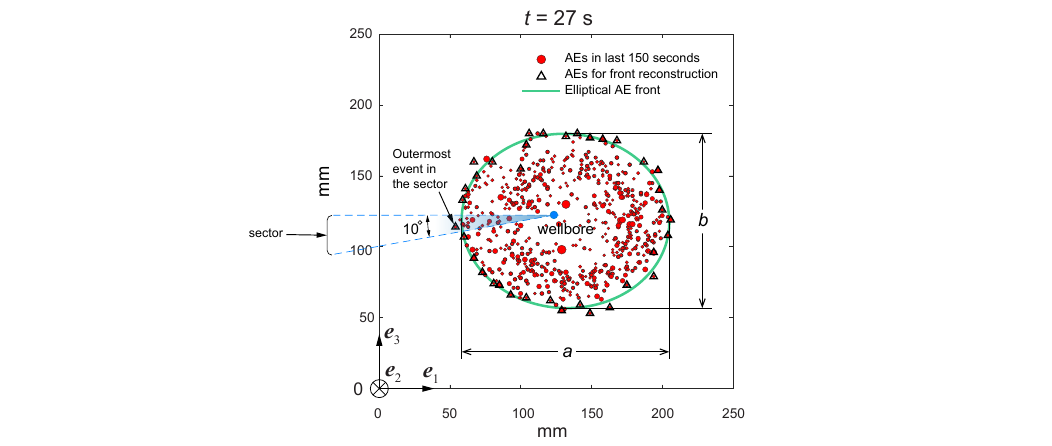}
\par\end{centering}
\caption{Illustration of the elliptical acoustic emission front reconstruction in test T2 ($t=27$ sec).\label{fig:T2_ellipse}}
\end{figure}

\begin{figure}
\begin{centering}
\includegraphics[width=\textwidth]{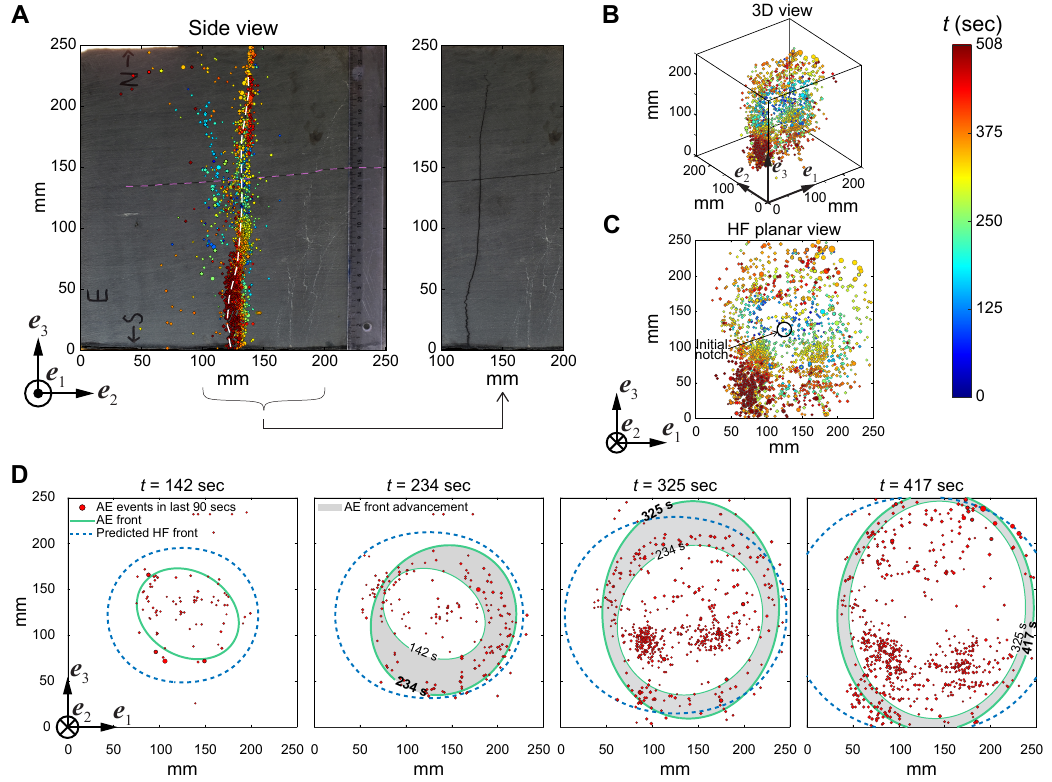}
\par\end{centering}
\caption{Experimental and simulation results of M4. The discrepancy between the acoustic emission contour and predicted fracture front is observed on the left side of the $\bm{e}_1\bm{e}_3$-plane. We noticed that acoustic emission data is largely missing in this region, which possibly is due to bad contact between the rock face and the piezoelectric sensors nearby. This could lead to error in the reconstruction of the acoustic emission front.\label{fig:M4_footprint}}
\end{figure}

\begin{figure}
\begin{centering}
\includegraphics[width=\textwidth]{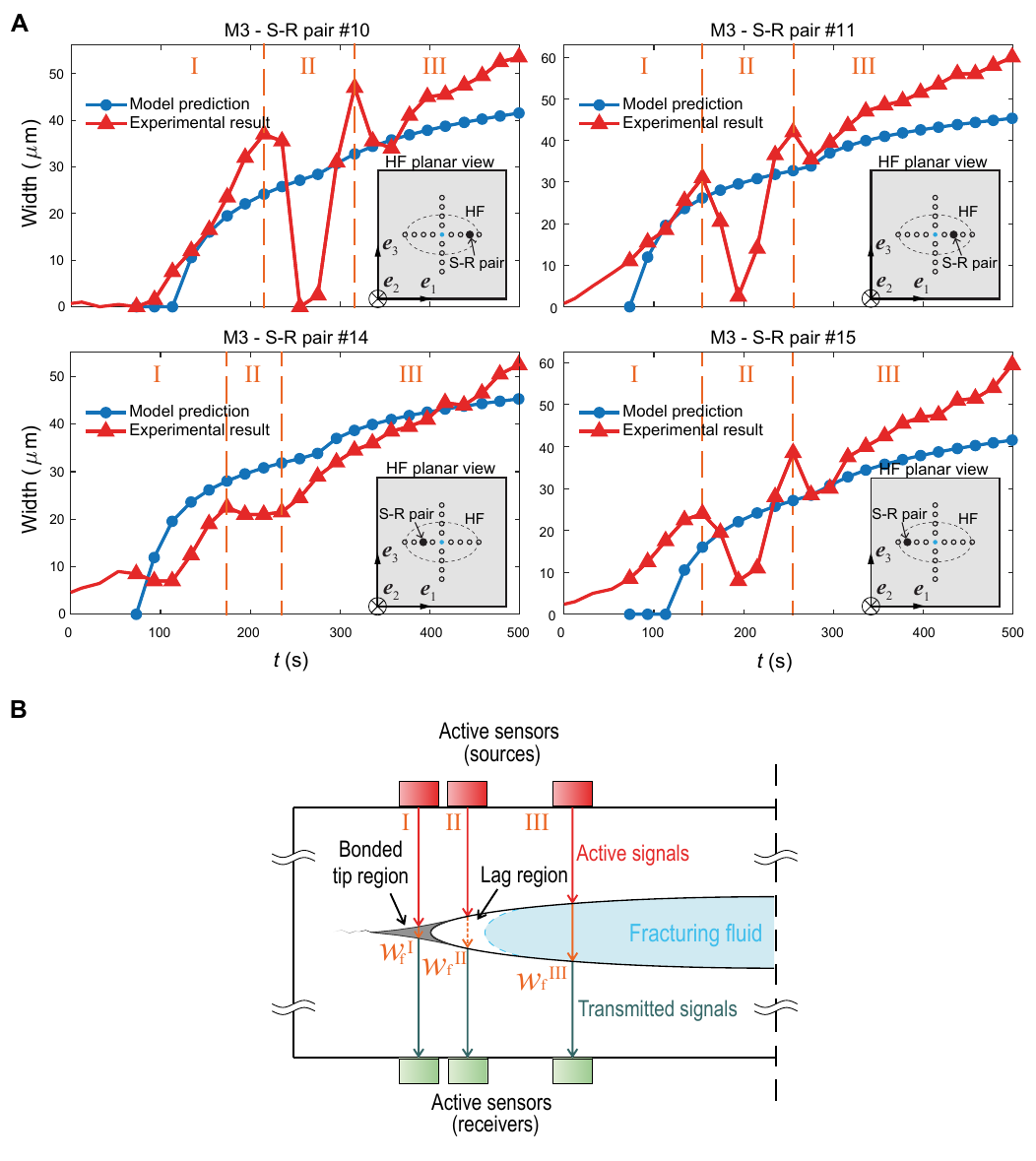}
\par\end{centering}
\caption{\textbf{A}, Width evolution at source-receiver pairs \#10, 11, 14 and 15 in M3. \textbf{B}, Conceptual sketch of the tip region undergoing three phases as fracture front crosses the compressional-wave pathway.\label{fig:width_evolution_M3}}
\end{figure}

\end{document}